\documentclass[journal]{IEEEtran}
\usepackage{multirow}
\usepackage{booktabs}
\usepackage[font=footnotesize]{subcaption}
\usepackage[font=footnotesize]{caption}
\usepackage{amsmath}
\usepackage{amssymb}
\usepackage{ragged2e}
\usepackage{float}
\usepackage{tikz}
\usepackage{cite}
\usepackage[T1]{fontenc}
\usepackage[nolist]{acronym}
\usepackage[linesnumbered,ruled,lined,hanginginout]{algorithm2e}
\usepackage{algpseudocode}

\usepackage{color}
\usetikzlibrary{patterns}

\SetAlgoHangIndent{0em}
\newtheorem{definition}{Definition}
\let\endshdefinition\enddefinition
\def\enddefinition{\strut\hfill$\square$\endshdefinition}

\makeatletter
\g@addto@macro\normalsize{%
  \setlength\abovedisplayskip{3.5pt}
  \setlength\belowdisplayskip{3.5pt}
}

\begin{document}
\IEEEoverridecommandlockouts

\pagenumbering{gobble}

This work has been submitted to the IEEE for possible publication. Copyright may be transferred without notice, after which this version may no longer be accessible.
	\newpage

\title{Eliminating Phase Misalignments in Cell-Free Massive MIMO via Differential Transmission}

\author{Marx M. M. Freitas, Stefano Buzzi,~\IEEEmembership{Senior Member,~IEEE} and Giovanni Interdonato, \IEEEmembership{Member,~IEEE}

\thanks{This work was supported by the EU under the Italian National Recovery and Resilience Plan (NRRP) of NextGenerationEU, partnership on ``Telecommunications of the Future'' (PE00000001 - program ``RESTART''); specifically, M. M. M. Freitas was supported by Structural Project NTN, Cascade Call INFINITE, CUP D93C22000910001, while S. Buzzi and G. Interdonato were supported by Structural Project 6GWINET, Cascade Call SPARKS, CUP D43C22003080001.}

\thanks{The authors are with the Dept. of Electrical and Information Engineering (DIEI), University of Cassino and Southern Lazio, 03043 Cassino, Italy (e-mail:\{marxmiguelmiranda.defreitas; buzzi; giovanni.interdonato\}@unicas.it). S. Buzzi and G. Interdonato are also with the Consorzio Nazionale Interuniversitario per le Telecomunicazioni (CNIT), 43124 Parma, Italy. S. Buzzi is also with the Dipartimento di Elettronica, Informazione e Bioingegneria (DEIB), Politecnico di Milano, 20156 Milan, Italy.}

}
\markboth{}{}

\maketitle

\begin{acronym}
    \acro{5G}{fifth-generation}
	\acro{AP}{access point}
	\acro{AWGN}{additive white Gaussian noise}
	\acro{ASD}{Angular standard deviation}
	\acro{ABC}{Artificial Bee Colony}
	\acro{AoA}{angle of arrival}
    \acro{ADC}{analog-to-digital converter}
	\acro{B5G}{beyond 5G}
	\acro{BA}{Bat Algorithm}
	\acro{BS}{base station}
	\acro{BER}{bit-error rate}
	\acro{BPSK}{binary phase shift keying}
	\acro{CS}{Cuckoo Search}
	\acro{CAPEX}{capital expenditure}
	\acro{CF}{cell-free}
    \acro{CF-mMIMO}{cell-free massive multiple-input multiple-output}
	\acro{CPU}{central processing unit}
	\acro{CC}{computational complexity}
    \acro{CM}{complex multiplications}
	\acro{CTMC}{continuous-time Markov chain}
	\acro{CSI}{channel state information}
	\acro{CSIR}{CSI at the receiver}
	\acro{CSIT}{CSI at the transmitter}
	\acro{CDF}{cumulative distribution function}
	\acro{CHD}{channel hardening}
	\acro{DL}{downlink}
	\acro{DCC}{dynamic cooperation clustering}
	\acro{D-mMIMO}{distributed massive multiple-input multiple-output}
    \acro{DFT}{discrete Fourier transform}
    \acro{DSP}{digital signal processor}
    \acro{DPSK}{differential phase-shift keying}
    \acro{DSTBC}{differential space-time block coding }
	\acro{EE}{energy efficiency}
	\acro{FS}{fiber switch}
	\acro{FD}{full duplication}
        \acro{FFT}{fast Fourier transform}
	\acro{FVP}{favorable propagation}
	\acro{FPA}{flower pollination algorithm}
	\acro{FA}{Firefly Algorithm}
	\acro{GA}{Genetic Algorithm}
        \acro{GOPS}{giga operations per second}
	\acro{GWO}{Grey Wolf Algorithm}
 \acro{GPP}{general purpose processor}
	\acro{HCPP}{hard core point process}
	\acro{i.i.d.}{independent and identically distributed}
	\acro{InH-open}{Indoor Hotspot Open Office}
	\acro{IoT}{Internet of Things}
	\acro{IC}{inter-coordinated}
	\acro{LOS}{line-of-sight}
	\acro{LP-MMSE}{local partial MMSE}
	\acro{LSFD}{large‐scale fading decoding }
	\acro{LSFB}{Largest-large‐scale fading}
    \acro{LMMSE}{linear minimum mean square error} 
    \acro{ML}{maximum likelihood}
	\acro{MD}{matched-decision}	
	\acro{MCMC}{markov chain monte carlo}
	\acro{MIMO}{multiple-input multiple-output}
	\acro{MF}{matched filter}
	\acro{m-MIMO}{massive-multiple-input multiple-output}
	\acro{MTBF}{mean time between failures}
	\acro{MR}{maximum ratio}
	\acro{MOFPA}{Multiobjective Flowers Pollination Algorithm}
	\acro{MMSE}{minimum mean square error}
	\acro{NLOS}{non-line-of-sight}
    \acro{NCC}{network-centric}
	\acro{NMSE}{normalized mean square error}
	\acro{NOMA}{non-orthogonal multiple access}
	\acro{NP}{no protection}
	\acro{NS}{non-scalable}
	\acro{NF}{noise figure}
    \acro{NR}{new radio}
	\acro{OMA}{orthogonal multiple access}
	\acro{OFDM}{orthogonal frequency-division multiplexing}
	\acro{OPEX}{operational expenditure}
	\acro{PRBs}{physical resource blocks}
	\acro{PD}{partial duplication}
	\acro{P-MMSE}{partial MMSE}
	\acro{P-RZF}{partial regularized zero-forcing}
	\acro{PDF}{probability density function}
	\acro{PSO}{particle swarm optimization}
    \acro{PA}{proposed approach}
    \acro{PEP}{pairwise error probability}
	\acro{QAM}{quadrature amplitude modulation}
	\acro{QoS}{quality-of-service}
	\acro{RF}{radio frequency}
	\acro{RSMA}{rate-splitting multiple access}
	\acro{RMSE}{root-mean-square deviation}
	\acro{RV}{random variable}
	\acro{RS}{radio Stripes}
    \acro{RU}{radio unit}
	\acro{SB}{serial buse}
	\acro{SDMA}{space-division multiple acess}
	\acro{SE}{spectral efficiency}
	\acro{SFP} {small form-factor pluggable}
	\acro{SHR}{self-healing radio}
	\acro{SIC}{successive interference cancellation}
	\acro{SCF}{scalable cell-free}
	\acro{SINR}{signal-to-interference-plus-noise ratio}
	\acro{SNR}{signal-to-noise ratio}
	\acro{TCO}{total cost of ownership}
	\acro{TDD}{time-division duplex}
	\acro{TR}{Technical Report}
  	\acro{TRP}{transmission and reception point}
	\acro{UatF}{use-and-then-forget}
	\acro{UAV}{unmanned aerial vehicle}
	\acro{UE}{user equipment}
	\acro{UL}{uplink}
	\acro{UMi}{Urban Micro}
	\acro{UC}{user-centric}
    \acro{UCC}{User-centric clustering}
    \acro{ULA}{uniform linear array}
\end{acronym}

\begin{abstract}

This paper proposes two approaches for overcoming access points' phase misalignment effects in the downlink of  \ac{CF-mMIMO} systems. The first approach is based on the differential space-time block coding  technique, while the second one is based on the use of differential modulation schemes. Both approaches are shown to perform exceptionally well and to restore system performance in \ac{CF-mMIMO} systems where phase alignment at the access points for DL joint coherent transmission cannot be achieved.
\end{abstract}

\section{Introduction}

\Ac{UC} \acf{CF-mMIMO} systems are a highly credited deployment for next-generation wireless networks \cite{ngo2024ultradense}. By employing multiple \acp{AP} across the coverage area to serve the \acp{UE}, these systems can provide a more uniform \ac{SE}, resistance to link blockages, and a better coverage probability than cellular networks 
\cite{BookCFemil2021}.

In the \ac{DL} of \ac{CF-mMIMO} systems, the \acp{AP} transmitting to a specific \ac{UE} need to coordinate their transmissions to ensure that their signals reach the \ac{UE} antenna with perfect phase alignment, so that they can constructively add together. 
Achieving such alignment in practice poses significant challenges, as it demands flawless \ac{CSI} and a shared knowledge of the phase of the local independent oscillators used at the \acp{AP} for modulating information-carrying signals. Although the impact of the first factor is frequently discussed in the literature, the second, considerably more complex issue, is rarely addressed.

In this context, several approaches have emerged to assess and possibly improve the performance of \ac{CF-mMIMO} systems in the presence of phase misalignment. Paper \cite{AsynchronousCellFree} proposes a rate-splitting strategy to enhance the data rate in non-coherent \ac{DL} transmissions between the \acp{AP} by dividing the messages into common and private parts. In \cite{PM_NCT}, the energy efficiency of coherent and non-coherent transmissions is analyzed using successive convex approximation and the Dinkelbach algorithm, and it is argued that non-coherent transmissions can serve as an alternative to coherent transmissions when fronthaul links have limited capacity. The study \cite{d2024coherent} analyzes the effects of phase misalignment within the realm of non-terrestrial networks, highlighting their significance in this environment as well.
In \cite{PM_GTEL, PM_Larson}, a partially coherent framework is investigated. Specifically, in \cite{PM_GTEL}, the \acp{AP} were divided into non-overlapping clusters, each managed by a dedicated \ac{CPU}. The \acp{AP} within the same cluster are assumed to be perfectly phase-aligned. Hence, the \acp{AP} in the same cluster coherently transmitted the same data stream to the \ac{UE}, while \acp{AP} from different clusters transmitted distinct data streams. Paper \cite{PM_Larson} extends the approach in \cite{PM_GTEL} by introducing an algorithm that groups \acp{AP} into clusters of phase-aligned \acp{AP}.  
Unlike \cite{PM_GTEL}, \cite{PM_Larson} does not rely on a very optimistic assumption of perfectly phase-aligned \acp{AP}. Despite the potential of these strategies to improve the performance of \ac{CF-mMIMO} systems in the presence of phase misalignment issues, none of them offered a final methodology to overcome these effects in \ac{CF-mMIMO} \ac{DL} transmissions. In addition, these approaches are based on the accuracy of the channel estimates, which may be ineffective in highly loaded scenarios with strong pilot contamination.

In this letter, we propose two novel approaches to mitigate the phase misalignment drawback in \ac{DL} \ac{UC} \ac{CF-mMIMO} systems. For this purpose, the well-known \ac{DSTBC} and \ac{DPSK} schemes are utilized. These techniques do not require channel knowledge at the receiver to detect the transmitted data; moreover, they are \textit{totally immune} to phase misalignment effects during data detection, offering a promising alternative to \ac{DL} coherent transmissions when \acp{AP} phase misalignments are not compensated for. Numerical results will show the effectiveness of the proposed approaches, both in terms of \ac{BER} and system throughput.

This letter is organized as follows. The next section contains the description of the considered system model and the mathematical model of the signal received at the \ac{UE} in the presence of phase misalignments. Section III contains a brief review of \ac{DSTBC}, while Section IV includes the description of the two proposed approaches to overcome \acp{AP} phase misalignments in the \ac{DL}. Numerical results are discussed in Section IV, while, finally, concluding remarks are provided in Section V.

\section{System Model}
\label{Sec:SystemModelNew}

We consider a \ac{CF-mMIMO} system consisting of $L$ \acp{AP} and $K$ single-antenna \acp{UE}. The \acp{AP}, each equipped with $N$ antennas, are connected to a \ac{CPU} via error-free fronthaul links. The system operates on \ac{TDD} protocol and assumes reciprocity for the \ac{UL} and \ac{DL} channels. We focus on \ac{DL} transmissions and consider that the channel $\mathbf{h}_{k,l} \in \mathbb{C}^{N\times 1}$ between the \ac{AP} $l$ and \ac{UE} $k$ undergoes an independent correlated Rayleigh fading, being defined as\footnote{The notation $\mathcal{N}_{\mathbb{C}} \big (\mu,\sigma^{2} \big )$ stands for a complex Gaussian random variable with mean $\mu$ and variance $\sigma^{2}$, while $\mathbb{E} \left \{ \, \cdot \, \right \}$ denotes statistical expectation.} $\mathbf{h}_{k,l} \sim \mathcal{N}_\mathbb{C} (\mathbf{0}_N, \mathbf{R}_{k,l})$. Here, $\mathbf{R}_{k,l} = \mathbb{E} \{\mathbf{h}_{k,l} \mathbf{h}_{k,l}^\mathrm{H}\} \in \mathbb{C}^{N \times N}$ represents the statistical covariance matrix of $\mathbf{h}_{k,l}$.

\subsection{UL Training and Phase Misalignment}

Each coherence block consists of $\tau_c$ complex samples, which are divided into $\tau_{p}$ for \ac{UL} pilots and $\tau_{d}$ for \ac{DL} data. During the \ac{UL} training phase, all \acp{UE} send mutually orthogonal pilot signals to the \acp{AP} for channel estimation\footnote{Although \ac{DSTBC} does not require \ac{CSI} for data detection, channel estimation is performed to compute the transmit precoders, which are crucial for suppressing co-channel interference.}, with some \acp{UE} reusing them if $\tau_{p} < K$. Once the \acp{AP} receive the pilots, the channels are estimated using \ac{LMMSE} estimation \cite{BookCFemil2021}.

The presence of uncompensated phase misalignments at the \acp{AP} is modeled through an additional phase term that modifies the channel in an uncontrollable way. Otherwise stated, the true channel between \ac{UE} $k$ and \ac{AP} $l$ is expressed as
\begin{equation}
\mathbf{g}_{k,l} =  e^{j \theta_l} \mathbf{h}_{k,l},
\label{Eq:channelPhases}
\end{equation}
\noindent where $\theta_l$ represents the uncontrollable phase of the oscillator in the transmit chain at the $l$-th \ac{AP}. This phase may also include phase noise effects, and it is assumed to remain constant for at least the time required to transmit two consecutive codewords of the \ac{DSTBC}, i.e., for a few symbol intervals.

\subsection{AP Clustering and DL  Transmission}

In \ac{UC} \ac{CF-mMIMO} systems, the generic $k$-th \ac{UE} is served by a subset of \acp{AP}, called \ac{AP} cluster, denoted as $\mathcal{M}_k \subset \{1,\ldots,L\}$. The association policy between  \acp{UE} and \acp{AP} can be represented by the binary variates $a_{k,l}$, for all $k=1, \ldots, K$ and $l=1, \ldots, L$; specifically, $a_{k,l}=1$ if $l \in \mathcal{M}_{k}$, and equals $0$ otherwise. All the \acp{AP} in $\mathcal{M}_k$ concur to the transmission of the same data symbol to the \ac{UE} $k$. In this letter, in order to provide a fair comparison with the \ac{DPSK} scheme outlined in the following, we assume that a PSK modulation is used\footnote{The extension of the proposed approaches to rectangular constellations will be reported in a future work.}. Hence, $\sum_{i=1}^{K} a_{i,l} \mathbf{w}_{i,l} s^{p}_{i} \in \mathbb{C}^{N \times 1}$ represents the data signal sent by \ac{AP} $l$ at discrete time epoch $p$. The received signal at \ac{UE} $k$  can be expressed as
\begin{equation}
    \centering
    {y}^{p}_{k} = \sum_{l=1}^{L} a_{k,l}\mathbf{g}_{k,l}^{\mathrm{H}}  \mathbf{w}_{k,l} s^{p}_{k} + \sum_{l=1}^{L} \sum_{i=1, i\neq k}^{K} a_{i,l} \mathbf{g}_{k,l}^{\mathrm{H}}  \mathbf{w}_{i,l} s^{p}_{i} + n_{k}^p,
    \label{Eq:ReceivedSignalTraditionalCF}
\end{equation}
\noindent where $\mathbf{w}_{i,l} \in \mathbb{C}^{N \times 1}$ denotes the precoding vector, which satisfies $\mathbb{E} \big\{ \left\| \mathbf{w}_{i,l} \right\|^{2} \big \} = \rho_{i,l}$, with $\rho_{i,l}$ being the fraction of power allocated to the \ac{UE} $i$ regarding the \ac{AP} $l$. The term $s^{p}_{i}$ is the complex symbol intended for \ac{UE} $k$ at time instant $p$, where $p = \{1,\dots, \tau_d\}$. In addition, $n_{k}^p \sim \mathcal{N}_{\mathbb{C}}\big ( \mathrm{0}, \sigma_{\mathrm{dl}}^{2} \big )$ denotes the receiver noise at time instant $p$. The detection of $s^{p}_{k}$ given ${y}^{p}_{k}$ can be performed using the following \ac{ML} criterion
\begin{equation}
    \centering
    \hat{s}^{p}_{k} = \mathop{\mathrm{arg\,min}}_{s \in \mathcal{S}} \, \big| {\angle y}^{p}_{k}  - \angle s \big|^{2}.
    \label{Eq:ML_CellFree}
\end{equation}
\noindent where the operator $\angle \cdot$ denotes the phase.

\section{A primer on \ac{DSTBC}}

\ac{DSTBC} techniques were introduced in early '00  (see \cite{Larsson_Stoica_2003} for a concise yet useful treatise) for use in point-to-point MIMO systems, to enhance transmission diversity when the receiver does not have access to the \ac{CSI}. They represent the differential version of classical space-time codes that permit achieving diversity gain by leveraging the use of multiple antennas at the transmitter and (possibly) at the receiver. In the following, we briefly review useful concepts related to \ac{DSTBC}.

Let us consider a system with $N_t$ antennas at the transmitter and one antenna at the receiver, and let $P$ represent the number of symbol periods spanned by a space-time codeword. Let $\mathcal{S} = \{ s^{1}, s^{2}, \dots, s^{J} \}$ denote the set of $J$ complex symbols, drawn from a unitary constellation,  to be transmitted to the receiver in a given channel realization. Then, $\mathcal{S}$ is divided into smaller subsets, each containing $n_s$ symbols. 
Let us denote by $\mathcal{S}^{t} = \{ s^{(t-1)n_s +1}, \dots, s^{t n_s} \}$  the $t$-th subset of $\mathcal{S}$, and by $G = \lfloor \tau_d/P \rfloor$, where $\lfloor \cdot \rfloor$ is the floor operation, their number.

Each subset $\mathcal{S}^{t}$ is then mapped onto a code matrix $\mathbf{X}^{t} \in \mathbb{C}^{N_t \times P}$, with $t$ also serving as the index of the code matrix. For the \ac{DL} data decoding procedure to be feasible, $\mathbf{X}^{t}$ must be a unitary matrix, i.e., it has to satisfy the condition $(\mathbf{X}^{t})^{\mathrm{H}}\mathbf{X}^{t} = \mathbf{I}_{N_t}$. Another desirable property is orthogonality, which helps reduce the computational complexity of detection at the receiver. However, orthogonal STBC matrices exhibit only a semi-unitary structure. Specifically, they satisfy the relation $(\mathbf{X}^{t})^{\mathrm{H}} \mathbf{X}^{t} = \sum_{j=1}^{n_s} \big|s^{j}\big|^2 \mathbf{I}_{N_t}$. To make $\mathbf{X}^{t}$ unitary, $\mathbf{X}^{t}$ is typically multiplied by a factor of $1/\sqrt{n_s}$.

A widely known orthogonal code matrix in the literature is the Alamouti matrix, which encodes two symbols ($n_s = 2$) across two transmit antennas ($N_t = 2$) and two symbol periods ($P = 2$). The Alamouti matrix is expressed as 
\begin{equation}
\mathbf{X}^{t} =
\frac{1}{\sqrt{2}}
\begin{bmatrix}
s^{1} & \phantom{-}(s^{2})^{*} \\
s^{2} & -(s^{1})^{*}
\end{bmatrix},
\label{Eq:Alamouti}
\end{equation}
where $s^{1}$ and $s^{2}$ are 
symbols taken from a generic subset $\mathcal{S}^{t}$. For configurations with more transmit antennas, other orthogonal matrices exist; as an example, if $N_t = 4$, the following orthogonal code matrix can be employed
\begin{equation}
\mathbf{X}^{t} = 
\frac{1}{\sqrt{3}}
\begin{bmatrix}
\phantom{\:}s^{1} & \phantom{\::}0 & s^{2} & -(s^{3})^{*} \\ 
\phantom{\,,}0 & \phantom{\,,}s^{1} & \phantom{\::}(s^{3})^{*} & \phantom{\:::}(s^{2})^{*} \\
-(s^{2})^{*} & -(s^{3})^{*} & \phantom{\::}(s^{1})^{*} & \phantom{\::}0 \\
\phantom{\:::}(s^{3})^{*} & -(s^{2})^{*} & \phantom{\:}0 & \phantom{\:::}(s^{1})^{*}
\end{bmatrix}.
\label{Eq:codeMatrix_Nt_4}
\end{equation}
This code matrix sends $n_s = 3$ symbols over $P = 4$ symbol periods, thus having the code rate of $R = 3/4$.

Once the data symbols are mapped onto $\mathbf{X}^{t}$ for $t = \{1,\dots, G\}$, $\mathbf{X}^{t}$ is differentially encoded by forming a new matrix

\begin{equation}
    \centering
    \mathbf{C}^{t} = \mathbf{C}^{t-1} \mathbf{X}^{t},
    \label{Eq:EncodingMatrix}
\end{equation}
\noindent where $\mathbf{C}^{0} = \mathbf{I}_{N_{t}}$. Since $\mathbf{X}^{t} $ is unitary, it readily follows that $(\mathbf{C}^{t})^{\mathrm{H}} \mathbf{C}^{t} = \mathbf{I}_{N_{t}}$. If $\mathbf{C}^{t}$ is transmitted to a single-antenna \ac{UE} over $P$ consecutive time intervals, the received block signal can be expressed as the following $P$-dimensional row vector:
\begin{equation}
    \centering
 \mathbf{y}^{t} = \mathbf{h}^{\mathrm{H}} \mathbf{C}^{t-1} \mathbf{X}^{t} + \mathbf{n}^{t},
 \label{Eq:RecSig_singleAPUE}
\end{equation}
\noindent with $\mathbf{n}^{t} \sim \mathcal{N}_{\mathbb{C}}\big ( \mathbf{0}, \sigma_{\mathrm{dl}}^{2} \mathbf{I}_{N_{t}} \big )$ representing the receiver noise and $\mathbf{h}^{\mathrm{H}} \mathbf{C}^{t-1}$ denoting the term that is unknown at the receiver. Then, $\mathbf{X}^{t}$ can be decoded by using two consecutive received blocks, i.e., $\mathbf{y}^{t}$ and $\mathbf{y}^{t-1}$, where $\mathbf{y}^{t-1} = \mathbf{h}^{\mathrm{H}} \mathbf{C}^{t-1} + \mathbf{n}^{t-1}$. The \ac{ML} detection of $\mathbf{X}^{t}$ corresponds to compute $\hat{\mathbf{X}}^{t} = \mathop{\mathrm{arg\,min}}_{\mathbf{X} \in \mathcal{X}_{N}} \, \big\| \mathbf{y}^{t}\!\left( \mathbf{X} \right)^{\mathrm{H}} - \mathbf{y}^{t-1} \big\|^{2}$ or equivalently \cite{Larsson_Stoica_2003}
\begin{equation}
    \centering
    \hat{\mathbf{X}}^{t} = \mathop{\mathrm{arg\,max}}_{\mathbf{X} \in \mathcal{X}_{N}} \, \mathrm{ML}^{t},
    \label{Eq:DifferentialDecodingTraceReal}
\end{equation}
\noindent where $\mathcal{X}_{N}$ denotes the set of all possible code matrices for a given unitary constellation $\mathcal{S}$. The term $\mathrm{ML}^{t}$ is obtained as
\begin{equation}
    \centering
    \mathrm{ML}^{t} = \mathrm{tr} \left \{ \mathrm{Re} \left \{  \mathbf{X} \, \left (\mathbf{y}^{t} \right)^{\mathrm{H}} \mathbf{y}^{t-1} \right \} \right \}.
    \label{Eq:ML_k}
\end{equation}

\section{\ac{DL} Transmission Schemes Robust to Phase Misalignment}

This section presents the proposed approaches to mitigate phase misalignment effects in \ac{UC} \ac{CF-mMIMO} systems based on \ac{DSTBC} and \ac{DPSK} schemes. Specifically, it demonstrates how to adapt them to the \ac{CF-mMIMO} scenario.

\subsection{Differential STBC for \ac{CF-mMIMO} Systems}

\begin{figure}[!t]
\centering
\includegraphics[scale=0.20]{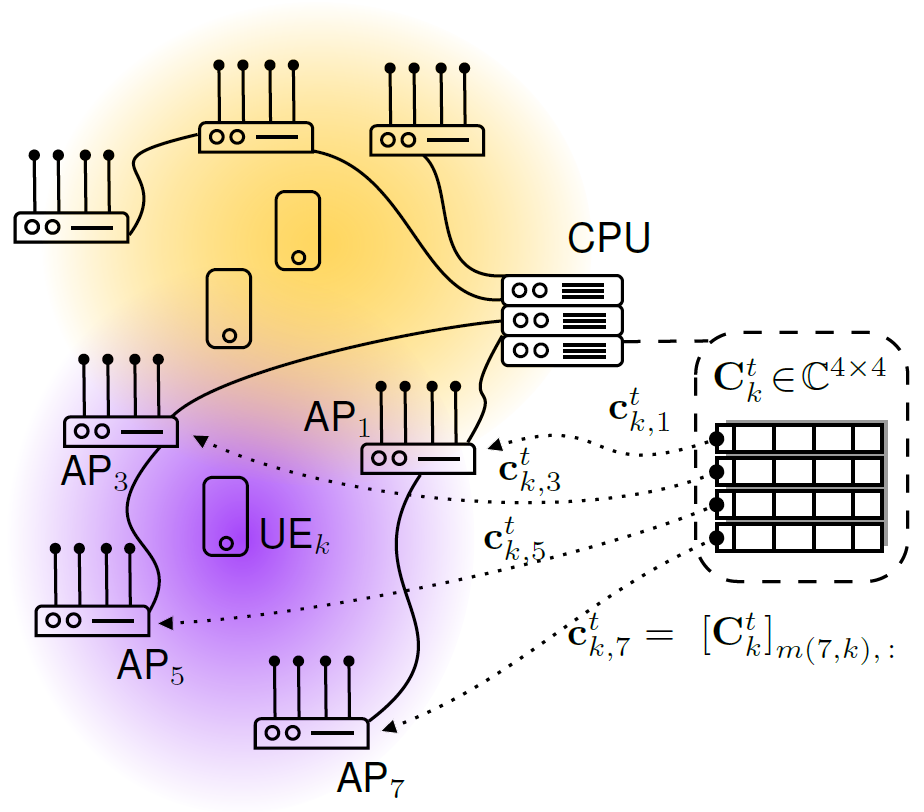}
  \caption{An illustration of how the information signal $\mathbf{C}^{t}_{k}$, designed at the CPU, is row-wise split among the \acp{AP} serving \ac{UE} $k$. Each row of matrix $\mathbf{C}^{t}_{k}$ is assigned to the corresponding \ac{AP} according to the mapping function $m(\cdot,k)$. In this example, the \acp{AP} transmit the elements of the assigned row over $L_k\!=\!4$ consecutive symbol intervals. As an example of the meaning of the mapping $m(l,k)$, note that, since the fourth row of the matrix $\mathbf{C}^{t}_{k}$ is sent to AP$_7$, we have that $m(7,k)=4$.}
  \vspace{-3mm}
  \label{Fig:splitC}
\end{figure}

To perform differential STBC techniques in \ac{CF-mMIMO}, we proceed as follows. 
First of all, a specific \ac{DSTBC} scheme is chosen, e.g., \eqref{Eq:Alamouti} or \eqref{Eq:codeMatrix_Nt_4}, even though many more choices are available \cite{Larsson_Stoica_2003}.
Next, each \ac{UE} must be connected to a number of \acp{AP} equal to the number of rows of the selected \ac{DSTBC} matrix. Then, with reference to \ac{UE} $k$,
the \ac{CPU} encodes $\mathbf{X}^{t}_{k}$ differentially as in \eqref{Eq:EncodingMatrix}. That is, it generates the information signal $\mathbf{C}^{t}_{k} \in \mathbb{C}^{L_{k} \times L_{k}}$ to be transmitted to the \ac{UE} $k$. Since the transmission involves multiple \acp{AP} instead of multiple antennas on a single transmitter, the \ac{CPU} then splits $\mathbf{C}^{t}_{k}$ among the \acp{AP} serving the \ac{UE} $k$, as depicted in Fig.\,\ref{Fig:splitC}. Specifically, each  row of the information matrix $\mathbf{C}^{t}_{k}$ will be assigned to one of the \acp{AP} serving \ac{UE} $k$; the elements of the rows of $\mathbf{C}^{t}_{k}$ will be transmitted by the associated \acp{AP} over $P$ consecutive time intervals. It is thus evident that in our proposed approach, each \ac{AP} acts like a single antenna of the original \ac{DSTBC}, and the multiple antennas at the \ac{AP} are used to suppress multi-user interference, not to perform the differential space-time coding task. 
The row of $\mathbf{C}^{t}_{k}$ that the \ac{CPU} assigns to \ac{AP} $l$ is computed as
\begin{equation}
    \centering
    \mathbf{c}^{t}_{k,l} = \big[ \mathbf{C}^{t-1}_{k} \big]_{m(l,k),:} \mathbf{X}^{t}_{k},
    \label{Eq:EncodingMatrixCellFree}
\end{equation}
\noindent where $l \in \mathcal{M}_{k}$ and $m(l,k) \in \left\{1, ..., L_{k} \right\}$ is a mapping function that 
points to the row of matrix $\mathbf{C}^{t}_{k}$ to be sent to \ac{AP} $l$ for transmission to \ac{UE} $k$.
Let $\sum_{k=1}^{K} a_{k,l}\mathbf{w}_{k,l} \mathbf{c}^{t}_{k,l} \in \mathbb{C}^{N \times L_{k}}$ represent the signal sent by \ac{AP} $l$. The received signal block at \ac{UE} $k$ can be expressed as
\begin{equation}
    \centering
 \mathbf{y}^{t}_{k} = \sum\limits_{l=1}^{L} \mathrm{g}_{k,l}^{\rm (ef)} \big[ \mathbf{C}^{t-1}_{k} \big]_{m(l,k),:} \mathbf{X}^{t}_{k} +  \widetilde{\mathbf{n}}_{k}^{t},
 \label{Eq:ReceivedSignalCellFreeSimplified}
\end{equation}
\noindent with $\mathrm{g}_{k,l}^{\rm (ef)} = a_{k,l}\mathbf{g}_{k,l}^{\mathrm{H}} \mathbf{w}_{k,l} \in \mathbb{C}$ representing the \textit{effective} \ac{DL} channel of \ac{UE} $k$. 
Moreover,
\begin{equation}
    \centering
     \widetilde{\mathbf{n}}_{k}^{t} = \sum_{i=1, i \neq k}^{K} \sum_{l=1}^{L} \widetilde{\mathrm{g}}_{i,k,l}^{\rm (ef)} \big[ \mathbf{C}^{t-1}_{i} \big]_{m(l,i),:} \mathbf{X}^{t}_{i} + \mathbf{n}^{t}_{k}
    \label{Eq:Noise_t}
\end{equation}
\noindent stands for the combined effects of interference and receiver noise, where $\widetilde{\mathrm{g}}_{i,k,l}^{\rm (ef)} = a_{i,l}\mathbf{g}_{k,l}^{\mathrm{H}}  \mathbf{w}_{i,l}$. One can note that \eqref{Eq:ReceivedSignalCellFreeSimplified} resembles \eqref{Eq:RecSig_singleAPUE}, as both assume that \ac{CSI} is not available at the receiver. Nevertheless, \eqref{Eq:ReceivedSignalCellFreeSimplified} provides a more general expression, as it is computed based on $\mathrm{g}_{k,l}^{\rm (ef)}$, which incorporates the effects of $\mathbf{h}_{k,l}$, the precoding, and phase misalignment.

To detect the code matrix $\mathbf{X}^{t}_{k}$, the \ac{UE} $k$ utilizes the \ac{ML} criterion defined in \eqref{Eq:DifferentialDecodingTraceReal}. However, the computation of $\mathrm{ML}^{t}_{k}$ in \eqref{Eq:ML_k} is based on the received signal model defined in \eqref{Eq:ReceivedSignalCellFreeSimplified}. To solve $\mathrm{ML}^{t}_{k}$, $\mathbf{y}^{t-1}_{k}$ is given by $\mathbf{y}^{t-1}_{k} = \sum_{l=1}^{L} \mathrm{g}_{k,l}^{\rm (ef)} \big[ \mathbf{C}^{t-1}_{k} \big]_{m(l,k),:} + \widetilde{\mathbf{n}}^{t-1}_{k},$ where $\widetilde{\mathbf{n}}^{t-1}_{k}$ is computed as $\widetilde{\mathbf{n}}^{t-1}_{k} = \sum_{i=1, i \neq k}^{K} \sum_{l=1}^{L} \widetilde{\mathrm{g}}_{i,k,l}^{\rm (ef)} \big[ \mathbf{C}^{t-1}_{i} \big]_{m(l,i),:} + \mathbf{n}^{t-1}_{k}$. Using the decision rule \eqref{Eq:ML_k}, the effects of phase misalignment are removed from the desired signal components of \ac{UE} $k$, as shown in the appendix.

\subsection{DPSK for \ac{CF-mMIMO}}
\ac{DL} coherent transmissions based on \ac{DPSK} schemes lie as a middle ground between conventional \ac{CF-mMIMO} systems and the above outlined \ac{DSTBC} schemes. That is, the information signal is transmitted individually at each time discrete time epoch $p$, for $p = { 1, \dots, \tau_d-1}$, and the received signal model follows \eqref{Eq:ReceivedSignalTraditionalCF}. However, the data symbols are differentially encoded. Let $\sum_{i=1}^{K} a_{i,l}\mathbf{w}_{i,l} \mathrm{c}^{p}_{i,l} \in \mathbb{C}^{N \times L_{k}}$ denote the signal transmitted by \ac{AP} $l$. The received signal at \ac{UE} $k$ can be modeled as
\begin{equation}
    \centering
    \mathrm{y}^{p}_{k} = \sum_{l=1}^{L} a_{k,l}\mathbf{g}_{k,l}^{\mathrm{H}}  \mathbf{w}_{k,l} c^{p}_{k} + \sum_{l=1}^{L} \sum_{i=1, i\neq k}^{K} a_{i,l}\mathbf{g}_{k,l}^{\mathrm{H}}  \mathbf{w}_{i,l} c^{p}_{i} + n_{k}^p,
    \label{Eq:ReceivedSignaDPSK}
\end{equation}
\noindent where $c^{p}_{k}$ is calculated as $c^{p}_{k} = c^{p-1}_{k} s^{p}_{k}$, and $c^{p-1} \in \mathbb{C}$ denotes the information signal transmitted at the previous time instant, with $c^{0}_{k} = 1$. Furthermore, since $s^{p}_{k}$ is taken from a unitary constellation, $\big( c^{p-1}_{k} \big)^{\mathrm{H}} c^{p-1}_{k} = 1$. The \ac{ML} detection of $s^{p}_{k}$ given $\mathrm{y}^{p}_{k}$ and $\mathrm{y}^{p-1}_{k}$ can be computed as \cite{Larsson_Stoica_2003}:
\begin{equation}
    \centering
    \hat{s}^{p}_{k} = \mathop{\mathrm{arg\,max}}_{s \in \mathcal{S}} \, \left \{\mathrm{Re} \{s\left( \mathrm{y}^{p}_{k}\right)^{*} \mathrm{y}^{p-1}_{k} \} \right \}.
    \label{Eq:ML_DPSK}
\end{equation}

\medskip
\noindent
\textit{Pros and Cons of the Proposed Approaches:}
The suggested methods enable \ac{DL} data decoding without requiring \ac{CSI} at the receiver. Specifically, \ac{CSI} estimates are utilized solely at the network side to calculate precoding vectors for the \acp{AP}. These approaches help mitigate phase misalignment in \ac{DL} transmissions. They are adaptable and compatible with any \ac{AP} selection strategy, with the only requirement being that each \ac{UE} must be served by a number of \acp{AP} matching the row count of the chosen \ac{DSTBC} matrix. Conversely, it's important to note that with the \ac{DPSK} strategy, the modulation is restricted to PSK, while in \ac{DSTBC}, the network's performance might be constrained by the code matrix rate\footnote{\ac{SE} results shown in the next section will permit evaluating the impact of the \ac{DSTBC} rate.}, expressed as the ratio $R = n_s/P$. For example, for $N_t > 4$, the code rates can significantly drop from $R = 3/4$. The Alamouti matrix shown in \eqref{Eq:Alamouti} is a particular instance with a code rate of 1 \cite{Larsson_Stoica_2003}. Lastly, data detection in \ac{DSTBC} demands more computation than in \ac{DPSK}, though, in the case of the here  considered \ac{DSTBC}, decisions on data symbols can be made independently, vastly simplifying the decision rule \eqref{Eq:DifferentialDecodingTraceReal}.

\section{Numerical Results}
Unless stated otherwise, we consider a \ac{CF-mMIMO} network consisting of $K = 20$ single-antenna \acp{UE} and $L = 40$ \acp{AP}, each equipped with $N = 4$ antennas. Moreover, each \ac{UE} is served by $L_{k} = 4$ \acp{AP}. The $K$ \acp{UE} are uniformly distributed over a square area of $0.5 \times 0.5$ km, while the positions of the \acp{AP} follow a \ac{HCPP}\footnote{This method ensures a minimum distance of $d_{\text{min}} = \sqrt{A/L}$ between any two \acp{AP}, where $A$ is the coverage area. Initially, the \acp{AP} are placed using a homogeneous Poisson point process with a mean rate of $1/d_{\text{min}}$. Then, the locations of any \acp{AP} that do not meet the spacing requirement are randomly updated until the condition is satisfied.}. The simulations focus on \ac{DL} channels, with the following parameters: $\tau_{c} = 200$, $\tau_{p} = 10$, and $\tau_{d} = 190$. The total transmission powers per \ac{UE} and \ac{AP} are set to $100\,\mathrm{mW}$ and $200\,\mathrm{mW}$. To compute the oscillator phases $\theta_l$, we utilize the discrete-time Wiener phase model \cite{MassiveMIMOPhases}.


To calculate the \ac{BER} and \ac{SE}, we perform Monte-Carlo simulations to account for different channel realizations and \ac{AP}/\ac{UE} locations, referred to as setups. For each setup, the received symbols (detected using the ML criterion) are mapped to bits using the Gray code, and the \ac{BER} is computed for each \ac{UE} over all channel realizations. The \ac{SE} for \ac{UE} $k$ in each setup is computed as:
\begin{equation}
    \centering
    \mathrm{SE}_k = P_f \log_2 \left( M_{o} \right) \left(1-\mathbb{E}\left\{ \mathrm{BER}_{k} \right \} \right)
\end{equation}
\noindent where $M_{o}$ is the modulation order and $\mathbb{E}\left\{ \mathrm{BER}_{k} \right \}$ is average BER across all channel realizations. It is assumed that the modulation order is $M_o = 8$. Moreover, $P_f$ is the pre-log factor, representing the fraction of samples in each coherence block dedicated to \ac{DL} data transmissions. $P_f$ is given by 
$\tau_d/\tau_c$ for conventional \ac{CF-mMIMO}, $(\tau_d-1)/\tau_c$ for \ac{CF-mMIMO} with DPSK, and $(G-1)n_s/\tau_c$ for \ac{CF-mMIMO}.
The \ac{AP} clustering scheme that jointly performs the pilot assignment and \ac{AP} selection is utilized \cite{BookCFemil2021}. In this scheme, each \ac{UE} is first associated with a coordination \ac{AP}. Then, the \acp{AP} serve only the \acp{UE} with the strongest channel gain in each pilot. The first $\tau_p$ \acp{UE} are assigned to orthogonal pilots, while the remaining ones are assigned to the pilot that causes the least contamination in each \ac{AP}. In the following, it is assumed that each \ac{UE} will be connected to the $N_t$ \acp{AP} with the strongest channel gains within its \ac{AP} cluster (or in its surroundings if $L_k < N_t$), subject to the constraint that no \ac{AP} is connected to more than $\tau_p$ \acp{UE}. We consider two types of processing: centralized and distributed. In the centralized one, the network processing (e.g., channel estimation and precoding vector generation) is centralized at the CPU. In the distributed one, the processing is distributed among the \acp{AP}.

The power coefficients at \ac{AP} $l$ in the distributed processing are given by $\rho_{k,l}=\rho_d {\sqrt{\beta_{k,l}}}/{\sum_{k^{\prime} \in  \mathcal{D}_l} \sqrt{\beta_{ k^{\prime},l}}}$, where $\rho_d$ is the maximum \ac{DL} transmit power per \ac{AP}. For the centralized one, the fractional power control is utilized \cite{BookCFemil2021}. To compute the precoding vectors, the \ac{P-MMSE} and \ac{LP-MMSE} are used for centralized and distributed processing, respectively \cite{BookCFemil2021}. The 3GPP \ac{UMi} path loss model is adopted for modeling the propagation channel \cite{3GPPSPEC}. It is considered that the shadowing terms of an \ac{AP} to different \acp{UE} are correlated, and the computation of correlation matrices $\mathbf{R}_{k,l}$ follows the local scattering spatial correlation model \cite{BookCFemil2021}. Table \ref{tab:UMiParameters} exhibits the parameters used in the models \cite{BookCFemil2021,3GPP5GNR}.

\begin{table}[!t]
    \renewcommand{\arraystretch}{1.1}
	\centering
	\caption{\scshape Simulation Settings}
	\label{tab:UMiParameters}
	\begin{tabular}{|c|c|}
    \hline
		\textbf{Parameter} & \textbf{Value} \\ \hline
		Shadow fading standard deviation, $\sigma_\mathrm{S F}$ & $4 \mathrm{~dB}$ \\ 
		\ac{AP}/\ac{UE} antenna height, $h_\mathrm{AP}, h_\mathrm{UE}$ & $11.65     \mathrm{~m}, 1.65 \mathrm{~m}$ \\
		RX \ac{NF} & $8 \mathrm{~dB}$ \\
		Carrier frequency, bandwidth & $3.5 \mathrm{GHz}, 20 \mathrm{MHz}$ \\
		\acp{ASD} & $\sigma_\varphi = \sigma_\theta = 15^{\circ}$\\
		Antenna spacing & 1/2 wavelength distance\\
    \hline
	\end{tabular}
    \vspace{-5mm}
\end{table}

\begin{figure*}[t]
    \centering
    \includegraphics[width=4.6cm,keepaspectratio=true]{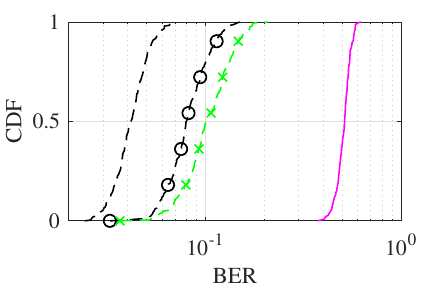} 
    \label{Fig:BER_LP_MMSE}
    \hspace{-0.47cm} 
    \includegraphics[width=4.6cm,keepaspectratio=true]{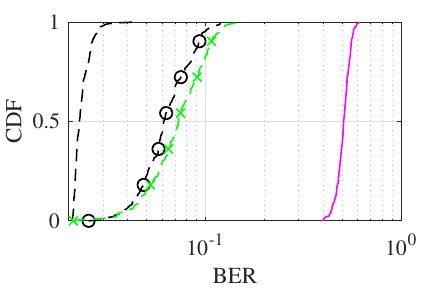} 
    \label{fig:P_MMSE}
    \hspace{-0.47cm}
    \includegraphics[width=4.6cm,keepaspectratio=true]{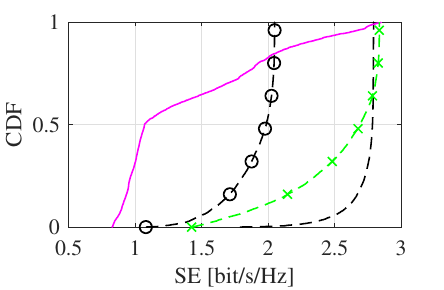} 
    \label{fig:SE_LP_MMSE_L_k_4}
    \hspace{-0.47cm}
    \includegraphics[width=4.6cm,keepaspectratio=true]{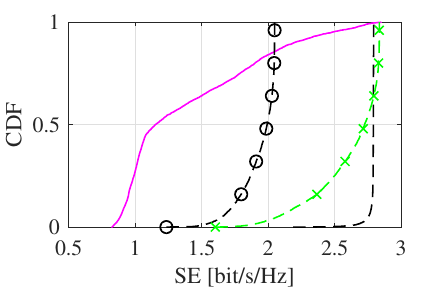} 
    \label{fig:SE_P_MMSE_L_k_4}
    \vspace{0.1cm}    
    \includegraphics[width=8cm,keepaspectratio=true]{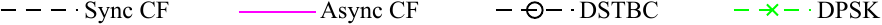}
    \vspace{0.125cm}  
    \begin{minipage}{\textwidth}
    \footnotesize
    \hspace{1.7cm}
    (a) LP-MMSE - BER \hspace{1.6cm} 
    (b) P-MMSE - BER\hspace{1.6cm} 
    (c) LP-MMSE - SE\hspace{1.7cm} 
    (d) P-MMSE - SE
\end{minipage}
    \caption{\Ac{CDF} of the average \ac{BER} and \ac{SE} for each setup. Parameters setting: $L = 40$, $K = 20$, $N = 4$, and $L_k = 4$.}
    \label{Fig:CDF}
    \vspace{-4mm}
\end{figure*}

Fig.\,\ref{Fig:CDF} presents the \acp{CDF} of the average \ac{BER} and \ac{SE} for \ac{CF-mMIMO} systems with (Async CF) and without phase misalignment (Sync CF). Both are compared with the proposed approaches. One can note that phase misalignment degrades the \ac{BER} of \ac{CF-mMIMO} systems performing \ac{DL} coherent transmission. However, this degradation can be mitigated by employing differential schemes, such as \ac{DSTBC} and \ac{DPSK}. This improvement is due to the ability of differential schemes to mitigate phase misalignment effects. Among these, \ac{CF-mMIMO} systems using \ac{DSTBC} achieve a lower \ac{BER} than those employing \ac{DPSK}, as \ac{DSTBC} not only mitigates phase misalignment, but also optimally  enhances transmission diversity.
\begin{figure}[htb!]
    \centering
    \begin{subfigure}{.25\textwidth}
    \centering
    \includegraphics[width=\textwidth]{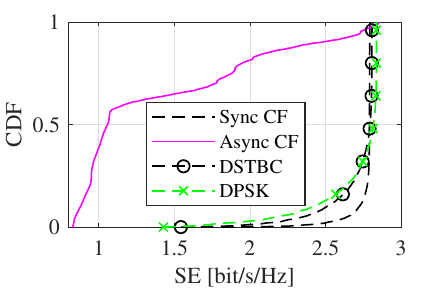} 
    \caption{\ac{LP-MMSE}}
    \label{Fig:CDF_SE_LP_MMSE_N_10_alamouti}
    \end{subfigure}
    \hspace{-0.5cm}
    \begin{subfigure}{.25\textwidth}
    \centering
    \includegraphics[width=\textwidth]{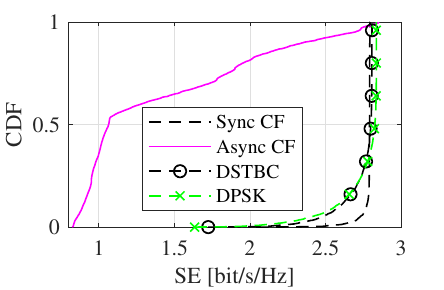} 
    \caption{\ac{P-MMSE}}
    \label{fig:CDF_SE_P_MMSE_N_10_alamouti}
    \end{subfigure}        
    \caption{\Ac{CDF} of the \ac{SE} for the proposed approaches and synchronized \ac{CF-mMIMO} system. Here, $L \!=\! 40$, $K \!=\! 20$, $N \!=\! 10$, and $L_k \!=\! 2$.}
    \label{Fig:CDF_SE_Alamouti}
    \vspace{-4mm}
\end{figure}

On the other hand, the \ac{SE} of \ac{DSTBC} is not as high as that of the \ac{DPSK} scheme. This difference is due to the pre-log factor of \ac{DSTBC} schemes, which is influenced by the code rate $R$, which is $R = 3/4$ for $L_k = N_t = 4$. Nonetheless, the code rate does not limit \ac{DSTBC} schemes for $L_k = 2$, since $R = 1$, as depicted in Fig.\,\ref{Fig:CDF_SE_Alamouti}. Moreover, the performance gap between \ac{CF-mMIMO} systems without phase misalignment and those employing differential schemes can be reduced if the \acp{AP} are equipped with more antennas, such as $N = 10$.

Finally, although not shown in this paper due to space constraints, our results confirm that the previous conclusions hold as the number of \acp{UE} $K$ increases. That is, differential schemes continue to mitigate phase misalignment effects. However, the \ac{SE} of \ac{DSTBC} degrades slightly more as $K$ increases. For instance, the \ac{SE} of the \ac{DSTBC} decreases by about 16.9\% for the 95\% likely \acp{UE} when $K$ increases from 20 to 40, while the \ac{SE} of \ac{DPSK} decreases by about 15\% for the \ac{P-MMSE} scheme. These results show that the \ac{DPSK} technique was slightly more resilient to interference effects than \ac{DSTBC} strategies in the considered scenario.

\section{Conclusions}
This paper proposed two novel approaches to mitigate phase misalignment effects in \ac{UC} \ac{CF-mMIMO} systems performing coherent \ac{DL} transmission. The first approach is based on the \ac{DSTBC} scheme, while the second relies on \ac{DPSK}. We demonstrated how to adapt these techniques to a \ac{CF-mMIMO} scenario and provided a mathematical proof of how \ac{DSTBC} overcomes phase misalignment in \ac{CF-mMIMO} systems. The proposed approaches were compared to a conventional \ac{UC} \ac{CF-mMIMO} network performing \ac{DL} coherent transmission. The simulation results revealed that the \ac{BER} and \ac{SE} of conventional \ac{CF-mMIMO} systems are severely degraded by phase misalignment, and that the proposed differential schemes are exceptionally effective in restoring performance. 

\appendices 

\section{Proof of Phase Misalignment Mitigation}
\label{Appendix_PM}

We now show how phase misalignment effects can be mitigated in a \ac{CF-mMIMO} system using \ac{DSTBC} for \ac{DL} transmission, based on the received signal model in \eqref{Eq:ReceivedSignalCellFreeSimplified}. By letting $\mathbf{DS}^{t}_{k} = \sum_{l=1}^{L} \mathrm{g}_{k,l}^{\rm (ef)} \big[ \mathbf{C}^{t-1}_{k} \big]_{m(l,k),:} \mathbf{X}^{t}_{k}$ denote the desired signal of \ac{UE} $k$, \eqref{Eq:ML_k} can be rewritten as
\begin{equation}
\centering
 \mathrm{ML}^{t}_{k}  =  \mathrm{Re} \left \{ \mathrm{tr} \left \{  \mathbf{X}_{k} \left( \mathbf{Y}_{k, 1} + \mathbf{Y}_{k, 2} + \mathbf{Y}_{k, 3} + \mathbf{Y}_{k, 4} \right )\right \} \right \},
 \label{Eq:ML_new}
\end{equation}
\noindent with $\mathbf{Y}_{k, 1} = (\mathbf{DS}_{k}^{t})^{\mathrm{H}}\mathbf{DS}^{t-1}_{k}$ denoting the desired signal components and the remaining terms corresponding to cross-products, calculated as $\mathbf{Y}_{k, 2} = (\mathbf{DS}_{k}^{t})^{\mathrm{H}} \widetilde{\mathbf{n}}^{t-1}_{k}$, $\mathbf{Y}_{k, 3} = ( \widetilde{\mathbf{n}}_{k}^{t})^{\mathrm{H}} \mathbf{DS}^{t-1}_{k}$, and $\mathbf{Y}_{k, 4}=( \widetilde{\mathbf{n}}_{k}^{t})^{\mathrm{H}} \widetilde{\mathbf{n}}^{t-1}_{k}$. Assuming that $\mathbf{X}^{t}_{k}$ is detected and since $\mathbf{X}^{t}_{k} (\mathbf{X}^{t}_{k})^{\mathrm{H}} =\mathbf{I}_{L_{k}}$, the product $\mathbf{X}^{t}_{k} \mathbf{Y}_{k, 1}$ in \eqref{Eq:ML_new} is calculated as
\begin{equation}
    \centering
     \sum_{l = 1}^{L} \sum_{l{'} = 1}^{L} \mathrm{g}_{k,l}^{{\rm (ef)}\,\mathrm{H}} \mathrm{g}_{k,l{'}}^{({\rm ef})} \left ( \big[ \mathbf{C}^{t-1}_{k} \big]_{m(l,k),:}\right)^{\mathrm{H}} \big[ \mathbf{C}^{t-1}_{k} \big]_{m{'}(l{'},i),:}.
    \label{Eq:X_k_Y1}
\end{equation}
\noindent Since the data symbols are taken from a unitary constellation, i.e., $|s| \!=\! 1$ for all $s \!\in\! \mathcal{S}$, the code matrices considered in this paper satisfy the property $\mathrm{tr} \big \{ \big(\big[ \mathbf{C}^{t-1}_{k} \big]_{m(l,k),:}\big )^{\mathrm{H}} \big[ \mathbf{C}^{t-1}_{k} \big]_{m(l,k),:} \big \} = 1$. Moreover, $\mathrm{tr} \big \{ \big(\big[ \mathbf{C}^{t-1}_{k} \big]_{m(l,k),:}\big )^{\mathrm{H}} \big[ \mathbf{C}^{t-1}_{k} \big]_{m{'}(l{'},i),:} \big \} = 0$ for $m{'} \neq m$ and $l \neq l{'}$. Consequently, $\mathrm{tr} \left \{\mathbf{X}^{t}_{k} \mathbf{Y}_{k, 1} \right \}$ is obtained as
\begin{equation}
    \centering
    \mathrm{tr} \left \{  \mathbf{X}^{t}_{k} \mathbf{Y}_{k, 1} \right \} = \sum_{l = 1}^{L} \mathrm{g}_{k,l}^{{\rm (ef)}\,\mathrm{H}} \mathrm{g}_{k,l}^{\rm (ef)} = \sum_{l = 1}^{L} \left | \mathrm{g}_{k,l}^{\rm (ef)} \right |^{2}.
    \label{Eq:trace_X_k_Y1}
\end{equation}
\noindent Therefore, although the interfering components (cross products) remain affected by phase misalignment, their impact on the desired signal is mitigated by the product $\mathrm{g}_{k,l}^{{\rm (ef)}\,\mathrm{H}} \mathrm{g}_{k,l}^{{\rm (ef)}}$.

\bibliographystyle{IEEEtran}
\bibliography{IEEEabrv,bib_file}

\vfill

\end{document}